\documentstyle[aps,prb]{revtex}

\begin{document}
\title{A study of photoexcited carrier relaxation in YBa$_{2}$Cu$_{3}$O$_{7-\delta
} $ by picosecond resonant Raman spectroscopy }
\author{T.Mertelj}
\address{Jozef Stefan Institute, Jamova 39, 1000 Ljubljana, Slovenia and\\
Faculty of Mathematics and Physics, Jadranska 19, 1000 Ljubljana, Slovenia}
\author{J. Demsar, B. Podobnik, I. Poberaj, D. Mihailovic}
\address{Jozef Stefan Institute, Jamova 39, 1000 Ljubljana, Slovenia}
\date{\today }
\maketitle

\begin{abstract}
The temperature dependence of the energy relaxation of photoexcited (PE)
carriers is used as a probe of the electronic structure of YBa$_{2}$Cu$_{3}$O%
$_{7-\delta }$ in the insulating ($\delta \approx 0.8)$ and metallic ($%
\delta \approx 0.1)$ phases. The energy relaxation rate to phonons is
obtained by measuring the non-equilibrium phonon occupation number, $n_{neq}$%
, with pulsed Raman Stokes/anti-Stokes spectroscopy using 1.5 and 70 ps long
laser pulses. We can distinguish between relaxation via extended band states
and localized states, since theoretically in the former, the relaxation is
expected to be virtually $T$-independent, while in the latter it is strongly 
$\ T$-dependent. From the experiment - which shows strong temperature
dependence of $n_{neq}$ - we deduce that at least part of the PE carrier
relaxation proceeds via hopping between localized states and we propose a
simple theoretical model of the relaxation process. In addition, we compare
the coupling of different vibrational modes to the carriers to find that the
apical O vibrational mode is significantly more involved in the energy
relaxation process that the in-plane 340 cm$^{-1}$ mode. This implies that
the localized states are mainly (but not entirely) coupled to out-of plane
vibrations.
\end{abstract}

\section{Introduction}

An important question in high-$T_{c}$ cuprates is whether the carriers that
lead to superconductivity in these materials can be described to be in
localized polaronic (magnetic or phononic) states or are in extended,
band-like states, albeit somewhat deviating from usual Fermi Liquid behavior
due to the shape of the Fermi surface. Since both views can be supported by
experimental observations\cite{Mott-Alexandrov}, either dual interpretations
are possible, whereby the experimental data can be explained in either
picture, or actually both types of carriers are present simultaneously. The
latter immediately leads to the possibility of a two component
superconductivity scenario, for example a Boson-Fermion scenario\cite
{Ranninger} or a 2+1 dimensional superconductivity scenario\cite
{Mihailovic-Egami} or even excitonic superconductivity\cite{Holcomb}.
Experimentally, the presence of localized states in the metallic phase of
the cuprates is suggested (amongst others) by photoconductivity measurements
on the insulating precursor YBa$_{2}$Cu$_{3}$O$_{7-\delta }$ (0.6$<\delta <$%
1.0)\cite{Yu91-92}, which show unambiguously that i) there are states within
2 eV of $E_{F}$ which are localized and ii) that at the insulator--to-metal
transition only the carriers near $E_{F}$ change their character giving rise
to a cross-over from semiconducting to insulating low-frequency conductivity 
$\sigma (\omega \rightarrow 0),$ while $\sigma (\omega )$ for $>$ 800 cm$%
^{-1}$ (0.1 eV) changes only gradually with doping and thus it is quite
likely that states further away from $E_{F}$ remain localized and Fermi
glass-like in the metallic phase.

In general, it is not possible to ascertain whether electronic states are
localized or extended using spectroscopy methods which measure the single
particle or joint density of states. More indirect methods are required,
especially for $\mid E-E_{F}\mid >kT$. By investigating the process of
carrier relaxation and energy loss from photoexcited states using Raman
spectroscopy, we present evidence of temperature-activated PE carrier
relaxation through localized states within $\sim $2 eV of $E_{F}$ in
insulating {\em and metallic }YBa$_{2}$Cu$_{3}$O$_{7-\delta }$ (with $\delta
\approx 0.8$ and $\delta \approx 0.1$ respectively).

To enable quantitative measurements of the phonon occupation numbers and
avoid possible misinterpretation of these results, we have employed
extensive calibration and correction procedures to quantitatively measure
the Raman Stokes and anti-Stokes intensities enabling an accurate
interpretation of the carrier relaxation process. These are described in
detail in the first Section of the paper. In the second Section, we present
the experimental results for different YBCO samples with a discussion of
calibration procedures and possible artifacts, and in the Discussion we
first discuss the connection between the carrier relaxation and the
observable quantities, then we suggest a model to explain the relaxation
process and finally discuss the origin of the localized states in the
context of the established knowledge about the electronic structure of the
material. The conclusions regarding the existence of localized states are
reached without assumptions about the mechanism for the electron-phonon
interaction and the PE process.

\section{Experimental technique and intensity calibration procedure}

The experimental technique used to measure the phonon shake-off entails the
measurement of the non-equilibrium phonon occupation number $n_{neq}$ by
measurement of the Stokes (S) and anti-Stokes (A) Raman scattering
intensities on a picosecond timescale. Details of the pulsed Raman
scattering experimental setup have been given previously\cite{Poberaj94}.
The experiment is performed by exciting carriers and simultaneously
measuring Raman scattered light with 1.5 ps or 70 ps laser pulses. These are
generated by frequency doubling a pulse-compressed (1.5 ps) or
non-compressed (100 ps) mode-locked Nd:YAG laser. The resulting photon
energy 2.33 eV (532 nm) is sufficient to excite inter-band transitions in
the material both within the CuO$_{2}$ planes and between the CuO$_{2}$
planes and the Cu-O chains depending on the polarization of the incident
light.\cite{Kircher91} The photon flux used is typically $10^{13}$ photons/cm%
$^{2}$/pulse, with estimated peak PE-carrier densities $n_{c}\approx 10^{18}$
cm$^{-3}$.

The measurements of $n_{neq}$ were performed on YBCO single crystals with $%
\delta \approx 0.1$ (twinned) and $\delta \approx 0.8$ mounted in a
liquid-He flow cryostat. The data are analyzed for the $A_{g}$ oxygen 340-cm$%
^{-1}$ plane-buckling vibration ($B_{1g}$ in tetragonal symmetry) and the $%
A_{g}$($A_{1g}$) 500-cm$^{-1}$ apical O(4) Raman-active vibrations. The
integrated intensities of the S and A phonon lines are obtained by fitting
to either a Gaussian or Lorentzian lineshape in the temperature range 100 -
320 K. The Gaussian lineshape was mainly used for the 340 cm$^{-1}$ mode,
where the frequency resolution of the spectrometer limited the linewidth.

The ratio of the Stokes and anti-Stokes Raman intensities in backscattering
geometry is given by: 
\begin{equation}
\frac{I_{S}\left( t\right) }{I_{A}\left( t\right) }=\frac{\xi \left( \omega
_{L\,},\omega _{S}\right) }{\xi \left( \omega _{L},\omega _{A}\right) }\frac{%
\omega _{S}\eta \left( \omega _{S}\right) }{\omega _{A}\eta \left( \omega
_{A}\right) }\frac{\Theta _{ab}\left( \omega _{L},\omega _{S},\omega
_{p}\right) }{\Theta _{ba}\left( \omega _{A},\omega _{L},\omega _{p}\right) }%
\frac{n_{p}\left( t\right) +1}{n_{p}\left( t\right) }.  \label{Is-Ia-ratio}
\end{equation}
Here $\omega _{p}$ is the phonon frequency, $n_{p}$ is the phonon occupation
number, $\hbar \omega _{L}$ is the incident-photon energy, $\hbar \omega
_{S} $ and $\hbar \omega _{A}$ the S and A scattered-photon energies
respectively, $\eta (\omega )$ is the frequency dependent refractive index
of the crystal and $\Theta _{ab}$ is the appropriate Stokes component of the
Raman tensor for incident polarization $a$ and scattered polarization $b$
while $\Theta _{ba}$ is the same for anti-Stokes scattering. We take into
account that due to time-reversal symmetry $\Theta _{ab}\left( \omega
_{L},\omega _{A},-\omega _{p}\right) =$ $\Theta _{ba}\left( \omega
_{A},\omega _{L},\omega _{p}\right) \,$\cite{Loudon63}. The factor $\xi
(\omega _{L},\omega )$ takes into account penetration depth and reflection
of incident and scattered light on the surface of the crystal at $\omega
_{L} $ and $\omega _{S}$ or $\omega _{A}$, as well as diffraction of the
scattered light upon exiting the crystal \cite{Cardona82}: 
\begin{equation}
\xi (\omega _{L},\omega )=\frac{\left( 1-\Re (\omega _{L})\right) \left(
1-\Re (\omega )\right) }{\kappa (\omega _{L})+\kappa (\omega )}\frac{1}{\eta
^{2}(\omega )}\text{,}  \label{corr-const}
\end{equation}
where $\Re (\omega )$ and $\kappa (\omega )$ are the reflectivity and the
absorption coefficients respectively.

Since it is known from experiment that the photoinduced reflectivity change $%
\Delta \Re /\Re $ $\sim 10^{-3}$ \thinspace for $\,$photoexcited carrier
densities\cite{Gong93} of $\sim 10^{21}$cm$^{-3}$, we assume that the change
in optical constants by photoexcited carriers with our carrier densities
typically of $\sim 10^{18}$cm$^{-3}$ is sufficiently small to ignore. Since
the relevant matrix elements for both the Raman and dielectric tensors are
related to each other, we assume that the same also holds for the Raman
tensor ${\bf \Theta }(\omega _{L},\omega _{s},\omega _{p})$.

The non-equilibrium phonon occupation number is defined by $n_{neq}\left(
t\right) =n_{p}\left( t\right) -n_{eq}\left( \omega _{p\,},T\right) $, where
at temperature $T$ for a phonon with frequency $\omega _{p}$, $n_{eq}$ is
given by:

\begin{equation}
n_{eq}\left( \omega _{p\,},T\right) =\left[ \exp \left( \frac{\hbar \omega
_{p}}{k_{B}T}\right) -1\right] ^{-1}\text{.}
\end{equation}
Since we are performing photoexcitation and Raman scattering with the same
laser pulse, the measured non-equilibrium phonon occupation number is a
weighted time average over the probe laser pulse duration: 
\begin{equation}
n_{neq}=\frac{\int g\left( t\right) n_{neq}\left( t\right) dt}{\int g\left(
t\right) dt}  \label{nneq-calc}
\end{equation}
where $g\left( t\right) $ represents temporal shape of the laser pulse.

To obtain accurate values of $n_{neq}$ one must take into account for the
fact that magnitudes of the Raman tensor components and the optical
constants are different for S and A photons due to the finite width of the
resonances with electronic states. Fortunately resonance Raman scattering
(RRS) has been extensively investigated in YBa$_{2}$Cu$_{3}$O$_{7-\delta }$%
\cite{Heyen90,Heyen92} and we use these data together with dielectric
constant data\cite{Kircher91} to calculate $n_{neq}$ from the measured S and
A intensities.

The spectrometer spectral response was calibrated by measuring the spectrum
of a tungsten lamp mounted in place of the sample. We assume that the lamp
spectrum is proportional to a black-body spectrum in the frequency range of
interest.\cite{Larrabee59} Since there was a spread in the ratio of the
spectral responses at $\omega _{S\text{ }}$and $\omega _{A}$ due to
spectrometer grating filling effects and lamp alignment, we averaged over
several calibration cycles to determine the spectrometer response. To check
our calibration procedure and whether the frequency dependence of the Raman
tensor and optical constants taken from the literature give the correct
value of $n_{p}$, the CW-laser intensity dependence of the phonon
temperature $T_{p}$ was determined from the $I_{S}/I_{A}$ ratio. The phonon
temperature is then given by 
\begin{equation}
T_{p}=\frac{\hbar \omega _{p}}{k_{B}\ln \left( \gamma _{p}\frac{I_{S}}{I_{A}}%
\right) }  \label{Phonon_T}
\end{equation}
where $\gamma _{p}\,$is the calibration constant to be determined for the
phonon $p$ including correction factors for the frequency dependence of the
spectrometer spectral response as well as the frequency dependence of the
Raman tensor and optical constants. To check whether $\gamma _{p}$ is
correctly determined, we measure $T_{p}$ (as determined using Eq. \ref
{Phonon_T}) using a CW laser as a function of laser power where under
stationary conditions, the phonon temperature is equal to the lattice
temperature in the scattering volume. We find that $T_{p}$ extrapolates to
the ambient temperature at zero incident laser power, confirming that our
calibration procedure is correct and giving an error in $T_{p}$ of less than
a 10 K.

The lattice temperature $T_{0}$ of the scattering-volume is somewhat higher
than the ambient cryostat temperature due to absorption of laser energy.
This needs to be corrected for if we wish to obtain an accurate value for $%
n_{p}(T)$ from our measurements. The temperature rise in the
scattering-volume can be divided into two parts. The first part, $\Delta
T_{CW}$ is a time-independent average rise of temperature which is a
consequence of heat buildup in the sample. This is very similar to the
heating due to absorption of a CW-laser beam with equivalent beam parameters
and average power, and depends on the sample geometry, the sample thermal
conductivities and thermal coupling of the sample to the sample holder.
Since all three components of the thermal conductivity tensor in YBCO are
virtually temperature independent down to 100K\cite{Bock95}, this is also
true for $\Delta T_{CW}$. We thus use the room-temperature value of $\Delta
T_{CW}$ measured with an Ar$^{+}$-ion CW laser with a photon energy of 2.41
eV (c.f. 2.33 eV for the Nd:YAG) as a function of laser power $P_{0}$ to
obtain an accurate calibration of $\Delta T_{CW}$ versus $P_{0}$.

The second correction is the transient temperature rise due to each pulse $%
\Delta T_{tr}$, and is determined mainly by the heat capacity of the
scattering volume. We calculate\cite{Bechtel85,MerteljThesis} that there is
only a $\sim 50\%$ increase in the transient part of $\Delta T$ for the 1.5
ps pulses in comparison to the 70 ps pulses, from which we deduce that most
of the generated heat remains in the scattering volume for the duration of
the laser pulse. $\Delta T_{tr}$ is calculated\cite{MerteljThesis} to be
below 0.5 K/mW for both 1.5 ps and 70 ps pulse lengths when focused to a 30 $%
\mu $m diameter spot. This is an order of magnitude smaller than the
observed phonon heating determined from the ratio of the S and A phonon
intensities at room temperature and is therefore neglected. Below 120 K, the
non-equilibrium effects we are trying to measure become comparable to the
effects of the time dependent part of the temperature increase $\Delta
T_{tr},$ and so we do not discuss data below this temperature.

\section{Experimental results}

\subsection{The nonequilibrium phonon temperature and nonequilibrium
occupation numbers}

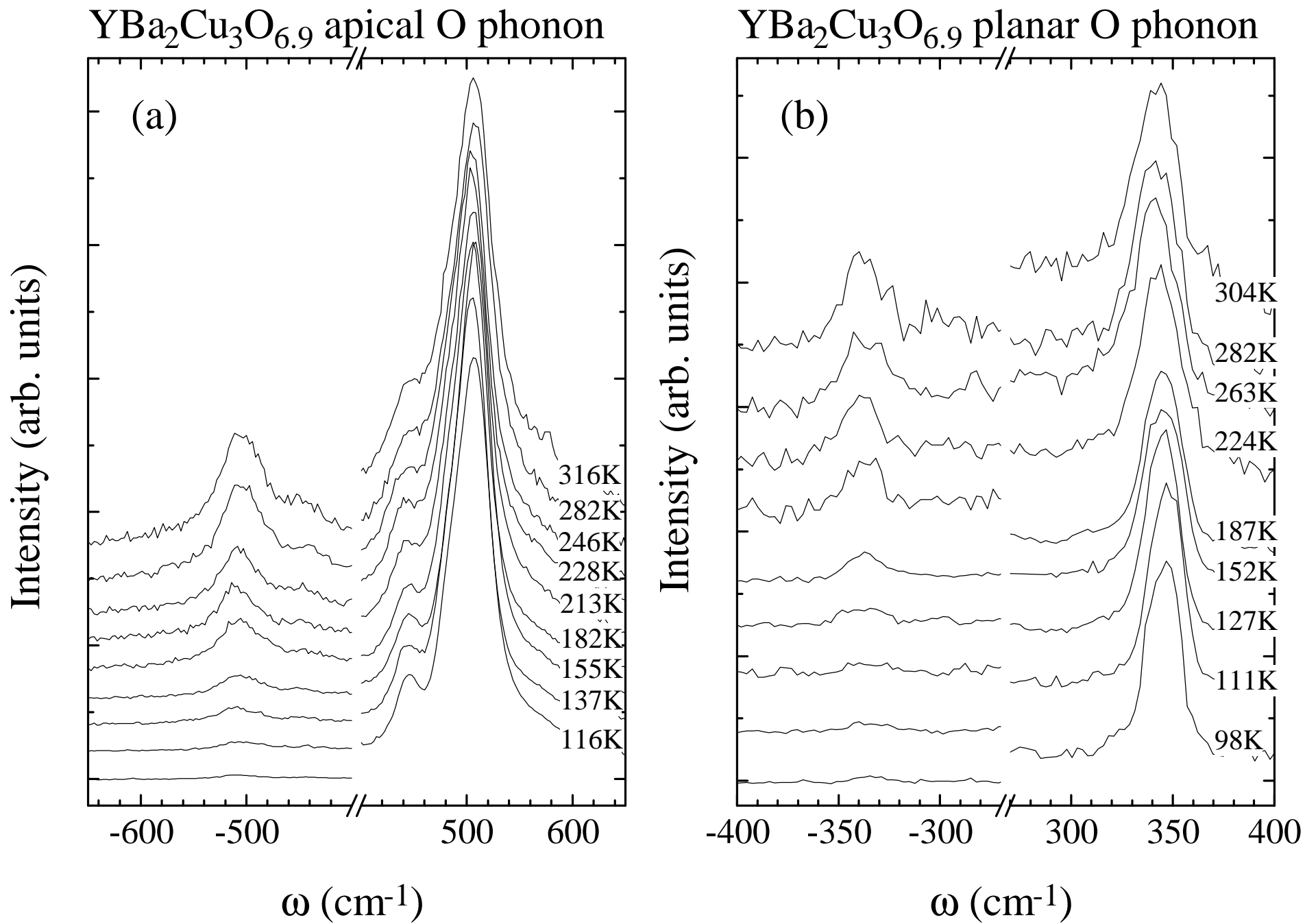
\begin{figure}
\caption{The phonon Raman spectra of the apical-oxygen phonon (a) 
and the planar buckling phonon (b) in $\delta \approx  0.1$ YBCO as 
a function of temperature at the 70-ps excitation. The spectra are vertically 
shifted for clearity. 
At lower temperatures the planar-buckling-phonon lineshape is 
distorted due to the low spectrometer resolution.}
\label{temp-spec-met}
\end{figure}

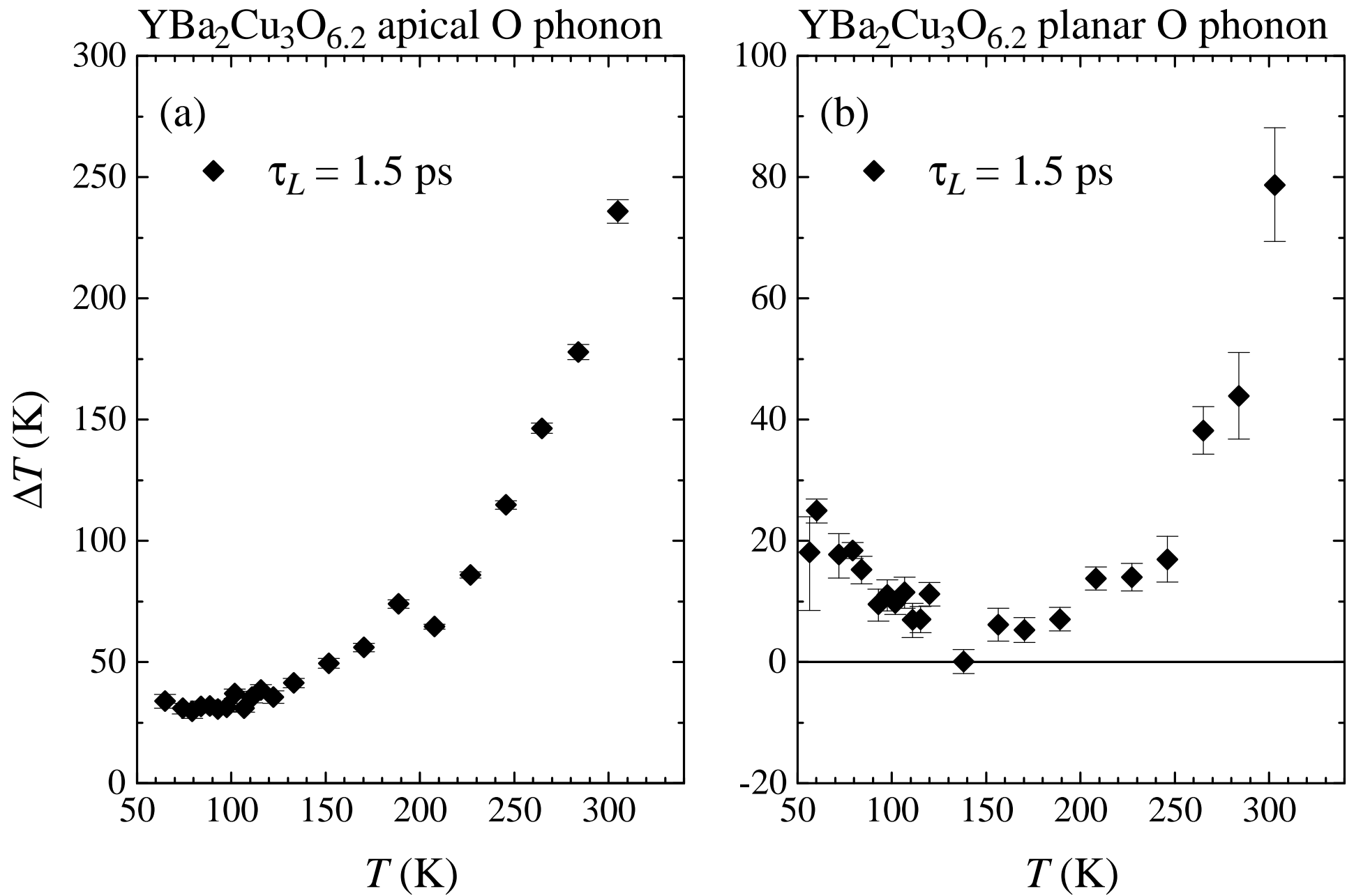
\begin{figure}
\caption{The nonequilibrium  phonon heating as a function of temperature for the 
apical-oxygen phonon (a) and the planar buckling phonon (b) in the 
$\delta \approx 0.8$ YBCO. Error bars represent standard errors obtained from fits of the 
Raman intensities.}
\label{dt-temp-ins}
\end{figure}

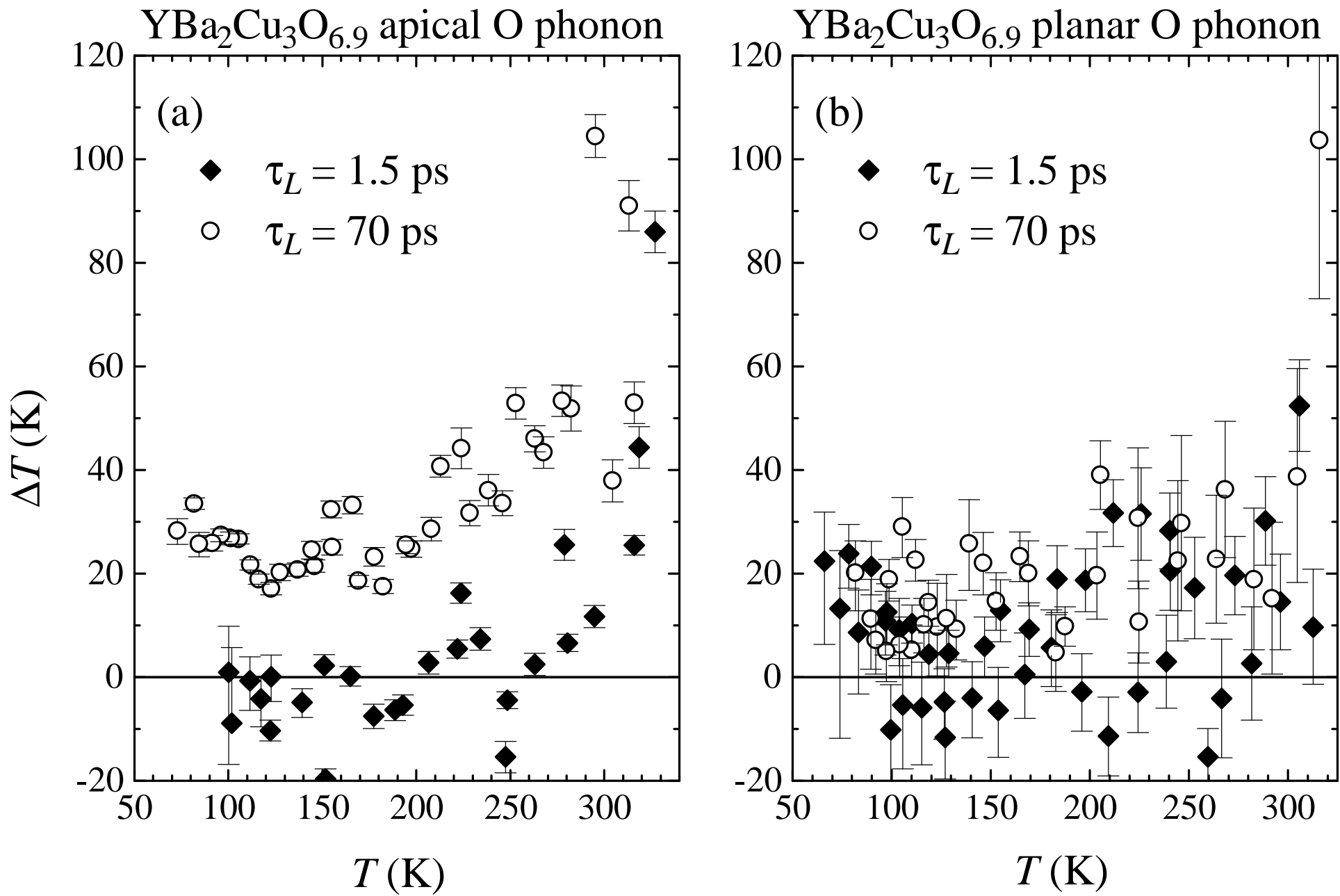
\begin{figure}
\caption{The nonequilibrium phonon heating as a function of temperature for the 
apical-oxygen phonon (a) and the planar buckling phonon (b) in the 
$\delta \approx 0.1$ YBCO. Error bars represent errors obtained from fits of the 
Raman intensities.}
\label{dt-temp-met}
\end{figure}

The normalized Raman spectra at different temperatures for YBCO with $\delta
\approx 0.1$\thinspace are shown in Fig. \ref{temp-spec-met} for the two
different polarizations showing the A and S spectra of the two phonons. We
see that the A peaks become smaller with decreasing temperature and
disappear below 100 K. The temperature dependence of the nonequilibrium
phonon temperature $\Delta T_{p}$, defined as $\Delta T_{p}=T_{p}-T_{0}$ is
plotted in Figs. \ref{dt-temp-ins} and \ref{dt-temp-met} for $\delta \approx
0.8$ and $\delta \approx 0.1\,$ respectively. In both cases panel a) shows
the apical 500 cm$^{-1}$ phonon and panel b) the 340 cm$^{-1}$ phonon values.

In the $\delta \approx 0.8$ sample (Fig.2), a significant nonequilibrium
phonon heating is observed {\it only} for the 1.5 ps-pulse excitation. The
nonequilibrium apical-oxygen phonon heating is the most significant,
reaching 250K at room temperature. With decreasing temperature, the heating
monotonically drops and it saturates at $\Delta T_{p}\sim 30$K below 100K.
On the other hand, the nonequilibrium planar-buckling-phonon heating is
smaller and it drops faster with decreasing temperature reaching a minimum
around 150 K. Below 100 K, $\Delta T_{p}$\thinspace rises again with
decreasing temperature because of the drop in the heat capacity below this
temperature results in a measurable transient increase in the lattice
temperature as discussed in the previous section.

In the $\delta \approx 0.1$ sample, similar nonequilibrium heating is
observed for both laser pulse lengths at room temperature(Fig. \ref
{dt-temp-met}). However, the actual values of $\Delta T_{p}$ are smaller
than in the $\delta \approx 0.8$ sample. Temperature dependence of the
nonequilibrium apical-oxygen phonon heating is different for the different
excitation-pulse lengths (Fig. \ref{dt-temp-met}a): with 1.5-ps pulse
excitation, $\Delta T_{p}\,\,$steeply drops with decreasing temperature and
reaches zero between 200-250K. The drop of $\Delta T_{p}$ with decreasing
temperature is less steep with 70-ps laser excitation and saturates at $\sim
25$K below 200K.

The planar 340 cm$^{-1}$ phonon in the $\delta \approx 0.1$ sample shows a
smaller $\Delta T_{p}$ than the apical-oxygen (Fig. \ref{dt-temp-met}b) over
the whole range of temperatures, but the difference between phonons is less
dramatic than in the $\delta \approx 0.8$ sample. The difference in behavior
at different pulse lengths is also less clear for the planar 340 cm$^{-1}$
phonon due to a larger scatter in the data, which is a consequence of the
smaller Raman intensity for this phonon at the $\delta \approx 0.1$ doping.

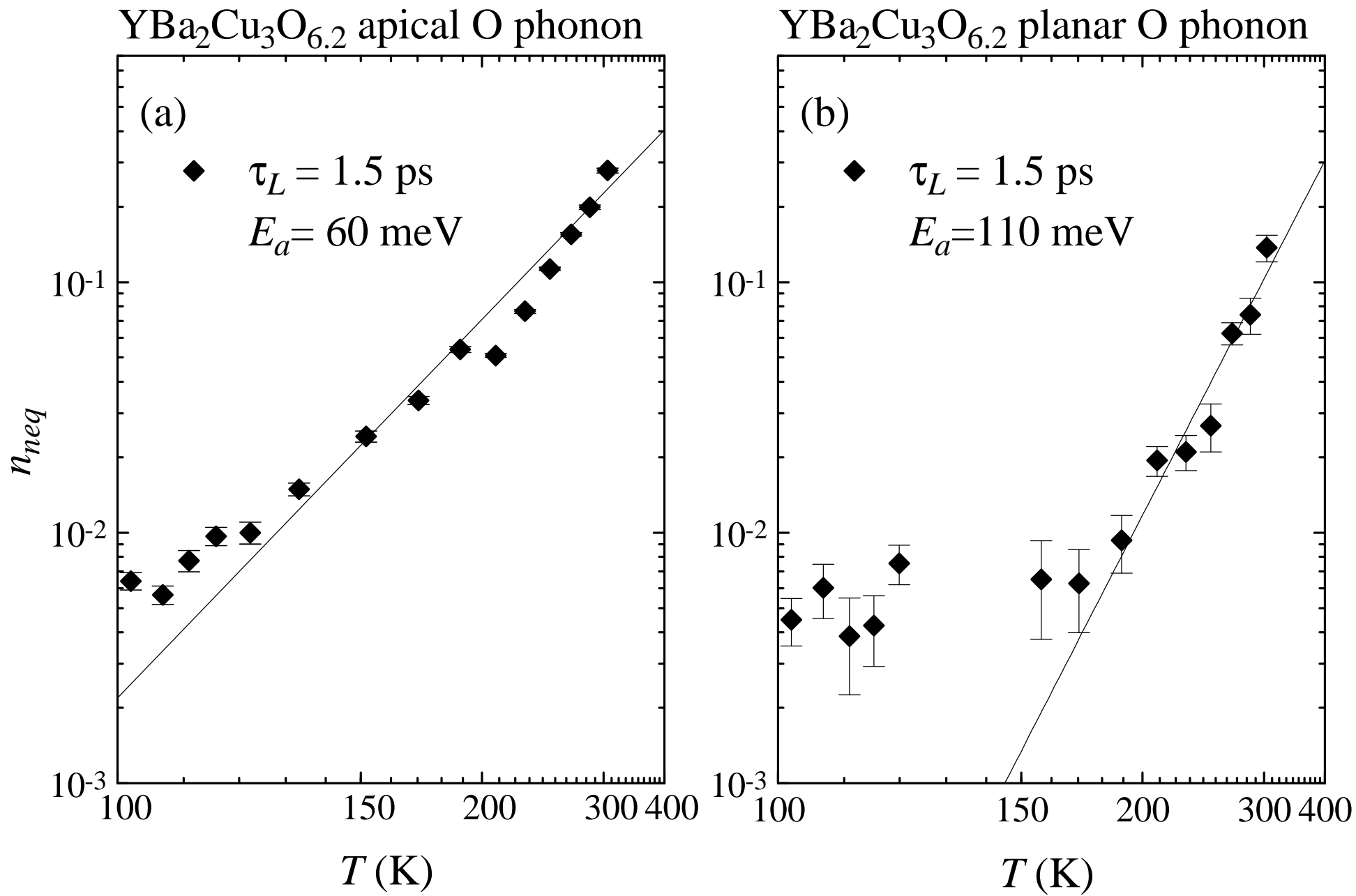
\begin{figure}
\caption{Arrhenius plots of $n_{neq}^{\delta=0.8}$ versus  
temperature for the apical-oxygen phonon (a), and the planar buckling phonon 
(b) in the $\delta \approx 0.8$ YBCO. The solid lines are Arrhenius fits to the data.
Error bars represent errors obtained from fits of S and A intensities.}
\label{nneq-temp-ins}
\end{figure}

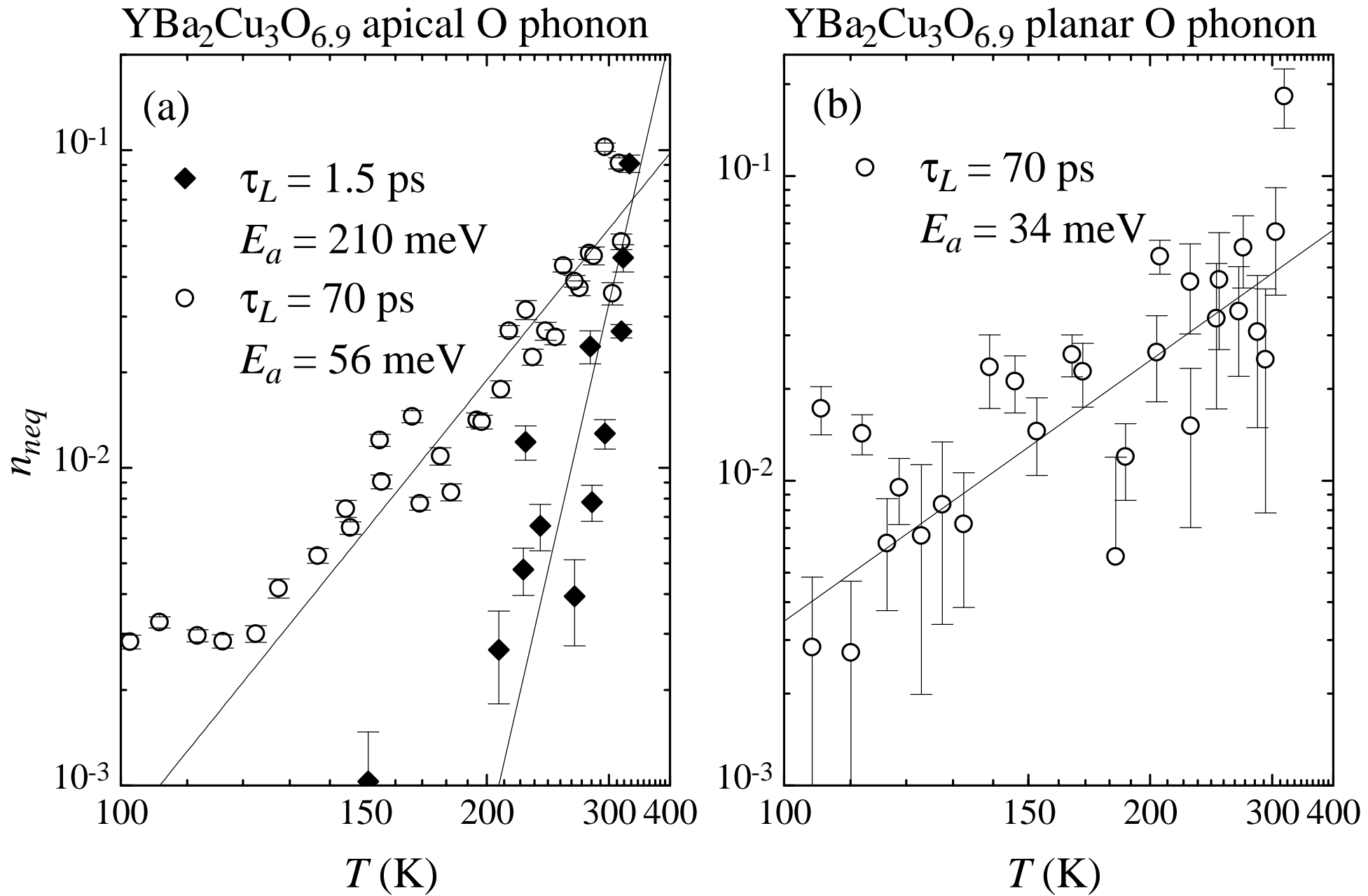
\begin{figure}
\caption{Arrhenius plots of $n_{neq}^{\delta=0.1}$ versus
temperature for the apical-oxygen phonon (a) and the planar buckling phonon (b) in 
the $\delta \approx 0.1$ YBCO. The solid lines are Arrhenius fits to the data.
Error bars represent errors obtained from fits of S and A intensities.}
\label{nneq-temp-met}
\end{figure}

Since the nonequilibrium-phonon occupation number $n_{neq}$ is related to
the carrier relaxation rate, its temperature dependence for both phonons is
plotted in Arrhenius plots in Fig. \ref{nneq-temp-ins} for $\delta \approx
0.8$ and Fig. \ref{nneq-temp-met} for the $\delta \approx 0.1$. The
important observation is that in all cases $n_{neq}$ shows a strong
temperature dependence. Fitting the data with an Arrhenius law, we obtain
activation energies $E_{a}$ in the range 56 - 210 meV in the metal ($\delta $
$\approx 0.1$) depending on pulse-length, and 60 - 110 meV in the insulating
phase ($\delta \approx 0.8)$. Below 120 K the transient laser heating
becomes comparable to the effect of $n_{neq}$, which is probably the reason
for the departure from the activated behavior (Fig. \ref{nneq-temp-met}) at
low temperatures. In the metal, for the apical O vibration, measurement with
both laser pulse lengths is possible and we observe a significant difference
in the activation energy in the two cases, which - as we shall see in the
Discussion - arises probably because of the effects of the temperature
dependence of the effective carrier lifetime.

\subsection{Discussion of possible artefacts in the Stokes/anti-Stokes ratios
}

In the measurement of Raman intensities, it is very important to take into
account all possible effects, to avoid erroneous interpretation of the
results. We thus proceed with a detailed analysis of the possible artefacts
with a detailed justification of all the assumptions made. For the
calculation of $n_{neq}$ from experimental S/A intensity ratios at different
temperatures it was assumed that correction factors and Raman tensor
components in (\ref{Is-Ia-ratio}) are temperature independent. These may
nevertheless have some, albeit small temperature dependence, and so the
approximation was checked on the basis of data available in the literature.
Since only the ratio of the correction factors and Raman tensor components
at the Stokes and anti-Stokes frequency appear in expression (\ref
{Is-Ia-ratio}), the frequency dependent part of the change of optical
constants and the Raman-tensor component with changing temperature has the
largest effect on the calculated $n_{neq}$, while the frequency independent
part of the change has only little effect.

For in-plane polarized light ($E||a,b$) Holcomb {\it et al.}\cite{Holcomb95}
measured the thermal difference reflectance in a $\delta \approx 0$ YBCO
thin film. They find that the quantity 
\begin{equation}
\Re (\omega ,T)^{-1}\frac{\partial \Re (\omega ,T)}{\partial T}\text{,}
\label{rel-ref}
\end{equation}
is below $4\cdot 10^{-4}$K$^{-1}$ around 2.3 eV photon energy in the 90-300K
temperature range and the difference of its values at $\omega _{S}$ and $%
\omega _{A}$ is below $\sim 10^{-4}$K$^{-1}$. By taking into account that
the reflectivity around this energy\cite{Kircher91} is 0.14 the total change
of reflectivity in this temperature range is estimated to be $\sim 0.012$,
giving relative change of $\sim 8\%$ with a difference at $\omega _{S}$ and $%
\omega _{A}$ of $\sim 2$\%. Because this is too small to be observed in our
measurements it is ignored.

Inferring further from the work of Fugol {\it et al.}\cite{Fugol93}, the
optical absorption in $\delta \approx 0.15$ YBCO film increases linearly by $%
\sim $2\% with increasing temperature in the temperature range 100K-140K.
Extrapolating to room temperature this would give a 10\% increase at 300K.
However, the increase is almost the same at 1.7eV and 2.7 eV indicating a
weak frequency dependence and hence this would have negligible effect on the
ratio of S and A intensities. The Raman-tensor component for the incident
and scattered light polarizations parallel to the CuO$_{2}$ planes decreases
with increasing temperature by $\sim 25\%$,\cite{Friedel91} but again the
decrease is virtually the same at 2.34eV and 2.6eV and is very similar at
1.92eV having no effect on the S/A ratio or $n_{neq}.$

The changes of the {\em frequency dependence} of optical constants and the
Raman-tensor component with temperature for light polarized parallel to the
CuO$_{2}$ planes is therefore estimated to be small enough that the
correction factors for the planar buckling-phonon can be considered
temperature independent in our experiment.

Regarding the c-axis polarized optical response, as far as we are aware
there are no published temperature dependence data for the visible region.
However, for $\delta \approx 0.1$ the low temperature resonant Raman
scattering data\cite{Heyen90} give the same value of the S/A intensity
correction factor at room temperature as measured by the CW Ar$^{+}$ laser
measurement, indicating that its temperature dependence is negligibly small.

For $\delta \approx 0.8,$ the experimental data on the temperature
dependence of optical response in visible region are also not available. The
data of Fugol {\it et al.}\cite{Fugol93} indicate that absorption for light
polarized parallel to the CuO$_{2}$ planes in a $\delta \approx 0.7$ YBCO
film is almost temperature independent at 2.7 eV while it shows temperature
dependence in the region of the 1.75-eV Cu-O charge-transfer absorption
peak. However, since this peak does not extend above 2eV, we expect that it
does not influence absorption at 2.33 eV significantly. In the absence of
other experimental data we expect that also for $\delta \approx 0.8$ YBCO
temperature dependence of the correction constants is small enough that it
does not significantly influence our interpretation of the results, while
the data for $\delta \approx 0.1$ are in principle more reliable, and the
assumptions are experimentally verified. In summary all the necessary
correction factors can be accurately obtained and cross-checked
experimentally giving us confidence of the accuracy of the measured $n_{neq}$%
.

\section{Discussion}

\subsection{Estimate of the carrier lifetime and discussion of the probing
process}

In a system as complex as YBCO there are many low energy excitations apart
from optical phonons (e.g. magnons) that can take up the relaxing-PE-carrier
energy. In our experiment, only the optical-phonon part of the relaxed
energy can be accessed. Despite unknown branching ratios to observed optical
phonons, the magnitude of the effective PE-cxarrier lifetime can be
estimated by comparing data for the 1.5 ps and 70 ps experiments in terms of
a simple model.

In the model we describe the PE carriers with their volume density, $n_{c}$,
and the effective lifetime $\tau _{c}$. The effective lifetime includes the
PE-carrier relaxation to all possible channels. The time-dependence of the
carrier number density $n_{c}$ is given by a simple relaxation-time
aproximation: 
\begin{equation}
\frac{dn_{c}(t)}{dt}=-\frac{n_{c}(t)}{\tau _{c}}+\kappa g(t)\text{,}
\label{pump}
\end{equation}
where $\kappa $ \thinspace is the absorption coefficient and $g(t)$ is the
incident photon flux density following the temporal profile of the laser
pulse. Assuming that at least a part of the PE-carrier energy is transferred
to optical phonons whose anharmonic lifetime is $\tau _{p}\,$(A part of the
PE carrier energy my be as well transfered to other low energy
excitations.), the time dependence of $n_{neq}$ is described by: 
\begin{equation}
\frac{dn_{neq}(t)}{dt}=-\frac{n_{neq}(t)}{\tau _{p}}+\frac{c_{pc}}{N_{p}}%
\frac{n_{c}(t)}{\tau _{c}}\text{.}  \label{phon-diff-eq}
\end{equation}
The second term on the right hand side is the phonon generation rate where $%
c_{pc}$ is the average number of phonons created by a carrier, which
implicitly includes also the branching ratio to optical phonons, and $N_{p}$
is the number of different phonon modes per volume unit involved in the
relaxation process.

Since a pump pulse is simultaneously a probe, the measured
nonequilibrium-phonon occupation number $n_{neq}$ is given by the equation (%
\ref{nneq-calc}). The laser pulse shape $g(t)$ is assumed to have a Gaussian
temporal profile, 
\begin{equation}
g(t)=\sqrt{\frac{2}{\pi }}j_{0}\frac{T}{\tau _{L}}\exp \left( -\frac{2t^{2}}{%
\tau _{L^{2}}}\right)  \label{pulse-shape}
\end{equation}
where $\tau _{L}$ is the laser pulse length, $T$ the pulse repetition period
and $j_{0}$ the average photon flux density.

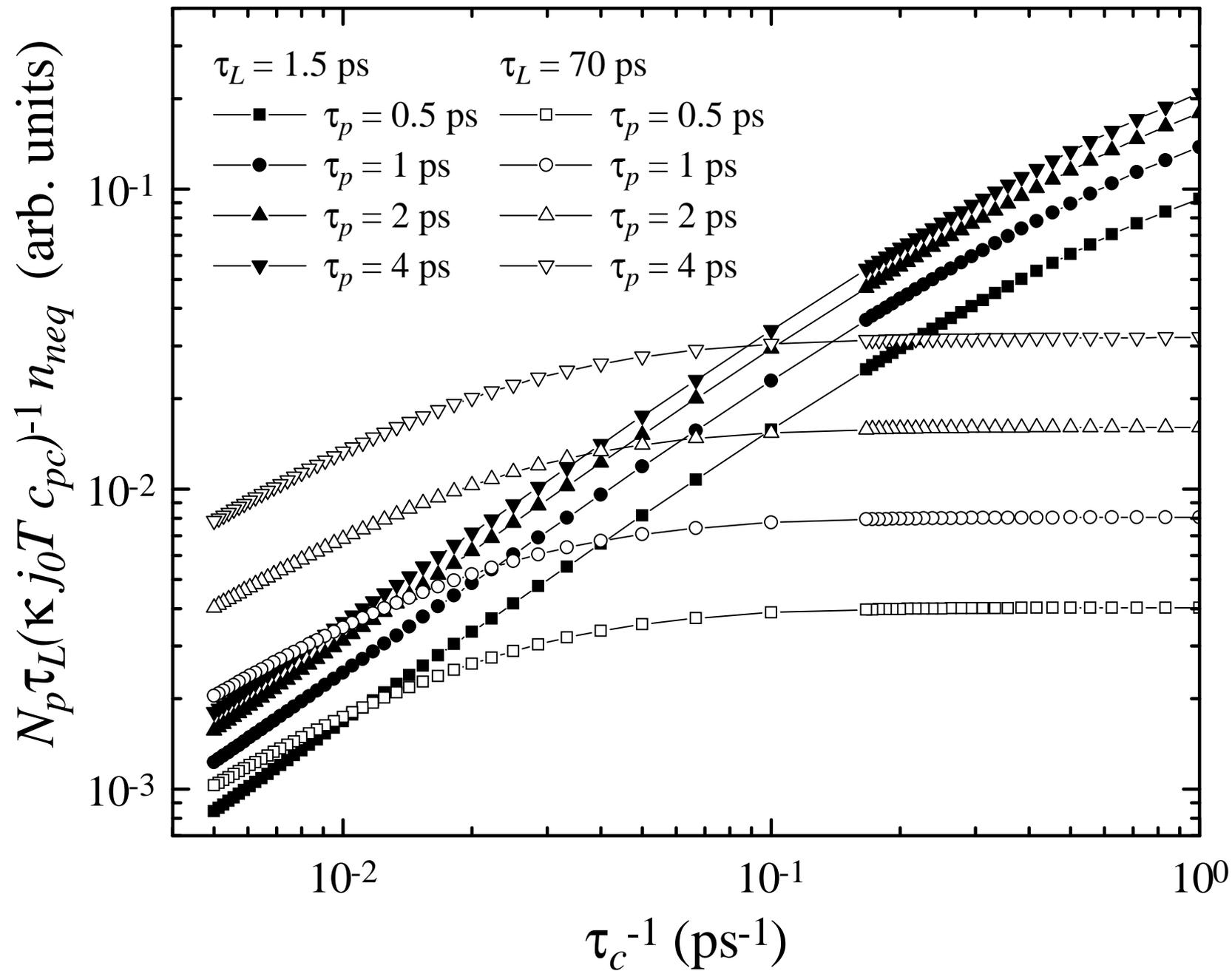
\begin{figure}
\caption{The calculated nonequilibrium phonon occupation number (\ref{nneq-calc})
as a function of the inverse PE-carrier relaxation time, $\tau _c^{-1}$, for two different pulse
lengths and different phonon lifetimes, $\tau_p$. The number of photons per pulse as wel as ratio $c_{pc}/N_p$ are 
the same for both pulse lengths. In this model $\tau_c^{-1}$ is proportional to the PE-carrier energy 
relaxation rate $r_E$.}
\label{nneq-calc-fig}
\end{figure}

Equations (\ref{pump}), (\ref{phon-diff-eq}) and (\ref{nneq-calc}) are
numerically integrated to obtain $n_{neq}$ as a function of the inverse
PE-carrier relaxation time, $\tau _{c}^{-1}$. The resulting curves for the
two experimental pulse lengths and different phonon lifetimes are shown in
Fig. \ref{nneq-calc-fig}. For 1.5-ps pulses $n_{neq}$ increases almost
linearly with increasing $\tau _{c}^{-1}$ for long PE-carrier lifetimes and
it saturates when $\tau _{c}^{-1}$ approaches $\tau _{L}^{-1}$. For the
70-ps pulses the initial increase is less steep, saturation is stronger due
to smaller value of $\tau _{L}^{-1}$ and it sets in at smaller values of $%
\tau _{c}^{-1}$. For both pulse lengths $n_{neq}$ increases with increasing
phonon lifetime with a stronger increase for the 70-ps pulses.

To enable comparison of the experimental data with the model, we estimate
the phonon lifetimes from the phonon Raman linewidths. At room temperature
for $\delta \approx 0.1,$ $\Delta \omega \approx 50$ cm$^{-1}$ and $\Delta
\omega \approx 20$ cm$^{-1}$ for the apical-oxygen phonon and the planar
buckling phonon respectively. For $\delta \approx 0.8,$ the apical-oxygen
phonon linewidth is $\Delta \omega \approx 30$ cm$^{-1}$\thinspace and the
planar-buckling phonon one is $\Delta \omega \approx 18$ cm$^{-1}$. Although
this includes also inhomogeneous broadening we can estimate the lower limit
for the phonon lifetimes which is sufficient for our discussion. We obtain $%
\tau _{p}\sim 0.2$ ps for the apical-oxygen phonon and $\tau _{p}\sim 0.4$
ps for the planar buckling phonon. This is consistent with the values of $%
\sim 0.5$ ps obtained for both phonons from experimental temperature
dependencies of phonon linewidths\cite{Mihailovic93,Friedel90,Mihailovic92}.

The room temperature value of $n_{neq}$ in the superconducting sample with $%
\delta \approx 0.1$ is nearly the same for both 1.5 and 70 ps pulse lengths.
Taking the room temperature phonon lifetime of $\sim 0.5$ ps we estimate
from the crossection of the curves for $\tau _{p}=0.5$ ps (the curves with
square symbols) in Fig. \ref{nneq-calc-fig} an effective PE-carrier-energy
lifetime to be in the range $10\sim 100$ ps. Here we use the fact, that the
position of the crossection along abscise does not depend on the unknown
coupling parameter $\frac{c_{pc}}{N_{p}}\,\,$and therefore on the branching
ratios to different excitations. We stress here that this is an estimate of
the time it takes for the carrier to reach equilibrium and not of its
lifetime in the excited state as implied by the model, which uses only a
simple relaxation-time aproximation for description of the PE-carier system.

Using the result that the phonon lifetime is apparently much smaller then
the effective PE-carrier lifetime, the number of emitted phonons per unit
volume is: 
\begin{equation}
N_{p}\approx \frac{c_{pc}}{n_{neq}}\frac{\tau _{p}}{\tau _{c}}n_{c}\text{.}
\label{number-of-phonons}
\end{equation}
In our experiment, the PE-carrier density $n_{c}\sim 4\times 10^{18}$ cm$%
^{-3}$. Since the photon energy is 2.33 eV and the phonon energy of $\sim 50$
meV, $10\sim 40$ phonons are released per absorbed photon. Taking the room
temperature experimental value of $n_{neq},$ we get $N_{p}$ $\sim $ 10$^{18}$
cm$^{-3}$ which is much smaller than the density of phonon modes in an
optical-phonon branch. Since Raman scattering probes the phonons with
wavevectors near the center of the Brillouin zone, this would imply that the
PE carriers relax primarily by emitting low momentum optical phonons near
the center of the Brillouin zone. Alternatively, in a scenario which is
consistent with the relaxation mechanism proposed in the next section -
together with the low dispersion and short anharmonic lifetime of the
optical phonons - the $k$-selection rule is no longer valid and the
vibrational modes excited by carriers are local modes for which the Raman $%
k\rightarrow 0$ selection rule is not relevant.

From Fig. \ref{nneq-calc-fig} it can be seen that $n_{neq}$ increases with
increasing phonon lifetime and decreases with decreasing PE-carrier
relaxation rate $\tau _{c}^{-1}$. Since $\tau _{p}$ {\em increases} slightly
with decreasing temperature, we can state with confidence that the observed
strong temperature dependence of $n_{neq}$ is not a result of the
temperature dependence of $\tau _{p}$. The temperature dependence of $%
n_{neq}\,$can as well be a consequence of temperature dependence of the
coupling parameter $\frac{c_{pc}}{N_{p}}$. In this case however, the
observed temperature dependence of $n_{neq}$ should be the same for
different excitation laser pulse lengths, which is not the case in our
experiment. We therefore conclude that that the observed strong temperature
dependence of $n_{neq}\,$\thinspace is mostly a consequence of a decrease of
the PE-carrier energy relaxation rate with decreasing temperature.

\subsection{Carrier relaxation via localized states}

The hot-carrier energy relaxation rate, $r_{E}$, in semiconductors with
hot-carrier energy $E>>kT$ and where carriers are assumed to relax via {\it %
extended} states, is virtually independent of the lattice temperature
irrespective of whether the electron-phonon interaction is via deformation
potential or via polar optical scattering.\cite{Seeger85}

In contrast, the measurements of $n_{neq}$ in YBCO show a strong temperature
dependence of $r_{E}$ for both $\delta \approx 0.1$ and $0.8$. In addition,
we find an unusually long room-temperature effective PE-carrier lifetime
(which is the total time the carrier is out of equilibrium) so we propose
that the PE-carrier relaxation in YBCO proceeds by {\it hopping through a
band of localized states.} There are however other possible mechanisms
giving temperature dependence of the PE carrier relaxation rate (e.g. strong
band dispersion or/and correlation effects). But the fact, that PE-carrier
relaxation virtually freezes out at low temperatures (as indicated by almost
two orders of magnitude drop of the $n_{neq}$ with decreasing temperature)
defenitely favors the PE carrier localization.

The temperature dependence of electronic relaxation in localized state
systems has so far been discussed in detail only in the context of charge
transport and the frequency-dependent conductivity $\sigma (\omega )$ \cite
{MottAndDavis79,Long82,Elliott87}. The relaxation processes discussed in
these works generally consider the hopping behavior of carriers near $E_{F}$%
, since they are interested primarily in low-frequency properties of such
materials. In our case the carriers are excited far above equilibrium and so
these models can not be directly applied, although the basic mechanisms for
hopping can be similar and indeed the predicted temperature dependence of
the conductivity is similar as we observe for the PE-carrier energy
relaxation rate.

The relaxation of PE\ carriers can proceed either via intra-gap defect
states as in a Fermi glass\cite{Yu91-92}, or through self-trapped polaron
states. The existence of polaronic self-trapped states in superconductor
insulator precursors with binding energies in the range of 60 - 100 mV has
been known from photoinduced absorption and fits to the mid-infrared region
(0.1-0.5 eV)\cite{polarons mid infrared}. The $T$-dependence of the hopping
rate in this case is essentially determined by the barrier height, $W$ which
is related to the polaron binding energy $E_{B}$ and so the temperature
dependence of $r_{E}$ is essentially given by the expression for activated
hopping\cite{MottAndDavis79}.

Alternatively in the Fermi-glass picture, the carriers relax to lower energy
mainly by tunneling through a cascade of neighboring states. However,
occasionally they have to hop upwards because there are no neighboring sites
with lower energies available giving rise to a temperature dependent
relaxation. The probability for transition upwards is proportional to $%
w_{0}\exp (-W/k_{B}T)$, where $W$ is the energy difference between
neighboring localized states, and $w_{0}$ the probability for transition
downwards. If $p$ is the probability that during relaxation a carrier hops
to a site with all neighboring sites at higher energy then the carrier
energy change after $N$ hops is 
\begin{equation}
\Delta E=WNp-WN\left( 1-p\right) =WN\left( 2p-1\right) \text{.}
\label{delta-e}
\end{equation}
The average time needed for N hops is 
\begin{equation}
\Delta t=Np\left[ w_{0}\exp \left( -\frac{W}{k_{B}T}\right) \right]
^{-1}+N(1-p)w_{0}^{-1}\text{,}  \label{delta-t}
\end{equation}
since the average time for a hop is inversely proportional to the transition
probability. The energy relaxation rate $r_{E}$ is then given by 
\begin{equation}
\frac{\Delta E}{\Delta t}=-\frac{w_{0}W}{p}\left[ \frac{1-p}{p}+\exp \left( 
\frac{W}{k_{B}T}\right) \right] ^{-1}\text{.}  \label{rate}
\end{equation}
At low enough temperatures, such that $k_{B}T<\frac{W}{\ln \left( \frac{1-p}{%
p}\right) }$, and neglecting the temperature-dependence of $w_{0}$, we
obtain:.

\[
\frac{\Delta E}{\Delta t}\approx -\frac{w_{0}W}{p}\exp \left( -\frac{W}{%
k_{B}T}\right) 
\]

To enable such behavior up to room temperature as observed in our
experiment, $p$ has to be larger than $\sim $0.2. If a uniform distribution
of localized states over an energy interval is assumed, then for a localized
state which lies in the middle of the energy band, the probability that all
its neighboring sites are at higher energy is 0.25 when each site has two
neighbors (chains) and 0.0625 in the case of four neighbors (planes), so our
experiments suggest that carriers are localized in an one-dimensional (1D)
structure.

For $\delta \approx 0.1,$ the chains are the most obvious 1D feature in
YBCO. However, apart of the chains there are also other possibilities of 1D
structures, for example a cross-section of twin boundaries and the CuO$%
_{2\,} $ planes, in-plane stripes\cite{Missori94} or 1D charge density wave
structures\cite{Edwards95} which have all been reported in YBCO. However,
the latter two are observed only at lower temperatures while the relaxation
scenario which we are discussing is essentially the same at room
temperature, apparently making these two possibilities less likely. In the
insulator, with $\delta \approx 0.8$ there are no fully occupied chains and
no twin boundaries, but the number of neighbors is small and there is some
local chain formation\cite{Andersen90} in the basal plane, so the picture is
also consistent with the proposed Fermi glass relaxation mechanism. The
activation energies we obtain from the fits in Figures 4 and 5 are
consistent with the values obtained in transient photoconductivity
measurements\cite{Yu91-92}, and also with polaron binding energies obtained
from fits to the mid-infrared conductivity spectra in YBCO\cite{polarons mid
infrared}. Although the value $E_{a}=210$ meV obtained for the apical O
vibration in the insulator is rather high, the implication of a deeper
potential for carriers trapped in the Cu-O chains is consistent with
calculations\cite{Ranninger1}.

\subsection{Discussion of the implications regarding the electronic structure
}

In our experiments the carriers are excited within 2.33 eV of the $E_{F}$.
To determine the final states of the PE process one has to identify the
relevant optical transitions at a given photon energy. Fortunately, for $%
\delta \approx 0.1$ YBCO, the optical transitions can be identified with
help of LDA\ band structure calculations together with experimental
ellipsometry data\cite{Kircher91}. For c-axis polarized light the
transitions are from the in-CuO$_{2}$-plane $pd\sigma $ antibonding bands
lying $\sim 1$ eV below the Fermi energy to the chain antibonding
O(4)-Cu(1)-O(1) $pd\sigma $ band\cite{Gopalan91}. For light polarized
parallel to the CuO$_{2}$ planes, transitions are largely (although not
exclusively) with both initial and final states in the CuO$_{2}\,$ planes.
The initial states lie $\sim 2$ eV below the Fermi energy, and the final
states lie just above $E_{F}$. In addition, there are also some transitions
between different Cu(1)-O(1) chain bands for light polarized along the $b$%
-axis.

In insulating YBCO with $\delta \approx 0.8$, optical transitions in this
energy range are somewhat harder to identify, since due to strong electron
correlation effects LDA calculations cannot be used as a reliable guide to
the transitions involved.\cite{Kircher91} However, from the resonant Raman
data on $\delta \approx 0.9$ samples Heyen {\it et al.}\cite{Heyen92}
conclude, that for light polarized parallel to the CuO$_{2}$ planes the
transitions involved in the energy range up to $\sim 2.5$ eV are from the
charge-transfer (CT) band to the upper Hubbard band (UHB). For c-axis
polarized light, the transitions in the same energy range are assigned\cite
{Heyen92} to transitions from initial states in the occupied Cu(1)$%
d_{3z^{2}-r^{2}}$-O(4)$p_{z}$ dumbbell band to final states in the UHB which
has also some Cu(2)$d_{3z^{2}-r^{2}}$ character.

Thus both in insulating and metallic YBCO, PE involves electronic
transitions from planes to chains (or Cu(1)-O(4) dumbbells in the case of
insulating YBCO) or vice versa for $c$-axis polarized light and mostly
transitions in the CuO$_{2}$ planes for in-plane polarized light. However,
since the PE carriers can partially relax before localization, the position
of localized states relative to the $E_{F}$ can not be inferred from the
position of the initial PE\ states, which is only the upper boundary. Taking
into account that in the present experiments only the PE carriers with
energies above the phonon energy are detected, the observed localized states
are at least one phonon energy away from the $E_{F}$. We cannot determine
whether the localized states are above or below $E_{F}$ since either holes
or/and electrons can relax producing optical phonons.

In insulating YBCO with $\delta \approx 0.8$ the apical oxygen phonon shows
significantly higher nonequilibrium phonon occupation numbers then the
planar-buckling one. This is consistent with stronger localization in the
basal plane or in the Cu(1)-O(4) dumbbells. On the other hand, for $\delta
\approx 0.1,$ the difference between planar-buckling and apical-oxygen
phonon nonequilibrium phonon effects is less dramatic, but somewhat
surprising is the implication by the data that localized states exist also
in the CuO$_{2}$ planes in the normal state of the optimally doped material.

The origin of the carrier localization is still not clear. Single-particle
spectra from angle-resolved photoemission spectroscopy\cite{photoemission}
imply the existence of a narrow band close to the Fermi energy. The origin
of the band narrowing could be Holstein narrowing due to lattice polarons
for example, or an effect of electronic correlations, or most likely both.
The bandwidth observed experimentally is comparable to $kT$ and lattice
imperfactions or defects can easily lead to carrier localization within such
a band. However, it is also apparent from ARPES\ that a - possibly
incoherent - background has non-zero spectral weight extending up to the
Fermi energy, and these states could well be responsible for the
temperature-dependent carrier relaxation. In any case, any tendency for
carrier self-localization effects will be enhanced even further by the
significant disorder present in these materials.

\section{Conclusions}

In conclusion, the pulsed resonant Raman scattering experiments show quite
clearly - and independently of the model for the relaxation process - that
some PE carrier relaxation proceeds via localized states. The comparison of
the measurements with 1.5-ps and 70-ps laser pulses indicates that in
metallic $(\delta \approx 0.1)$ YBCO the effective room temperature
PE-carrier lifetime before it reaches equilibrium is in the range of 10-100
ps, while for $\delta \approx 0.8$ the lifetime is shorter, probably in the
1 ps range. The present data and relaxation scenario are in agreement with
recent photoinduced absorption measurements by Thomas {\it et al.}\cite
{Thomas96} who observed a long-lived relaxation component after laser
photoexcitation in metallic YBa$_{2}$Cu$_{3}$O$_{6.9}$ and Bi$_{2}$Sr$_{2}$%
CaCu$_{2}$O$_{8+\delta }$ which they also attributed to carrier localization
effects in the normal state of the superconductor.

Theoretically, relaxation via either polaronic states or a Fermi glass
relaxation scenario can both reproduce the experimental temperature
dependence of the carrier relaxation, the latter model suggesting that the
states through which the relaxation takes place are in one-dimensional
chains. Although the position of the localized states in real space can be
discussed with some degree of certainty, we can only state that the
localized states lie within a 2.3 eV wide band around $E_{F}$ and relaxation
measurements - either Raman or time-resolved photoinduced absorption - with
different laser wavelengths would be necessary to determine their energy
more precisely. The presence of localized states in the optimally doped
superconductors is believed to have important consequences both for
superconductivity and hot-carrier device design using these materials.

%

\end{document}